\documentclass[twocolumn,showpacs,amssymb,aps]{revtex4}

\usepackage{graphicx}
\usepackage{psfig}
\usepackage{dcolumn}
\usepackage{bm}

\def\lessim{\lower.5ex\hbox{$\; \buildrel < \over \sim \;$}}
\begin{document} \hbadness=10000
\topmargin -0.8cm\oddsidemargin = -0.7cm\evensidemargin = -0.7cm
\preprint{}

\title{Centrality dependence of strangeness and (anti)hyperon production at BNL  RHIC}
\author{Jean Letessier}
\affiliation{Laboratoire de Physique Th\'eorique et Hautes Energies\\
Universit\'e Paris 7, 2 place Jussieu, F--75251 Cedex 05
}
\author{Johann Rafelski}
\affiliation{Department of Physics, University of Arizona, Tucson, Arizona, 85721, USA}

\date{June 14, 2005}

\begin{abstract}
We evaluate  strangeness produced  in Au--Au interactions at 
 $\sqrt{s_{\rm NN}}=200$ GeV, as function of reaction participant number $A$,
and obtain the relative strange quark content at hadronization.
Strange baryon and antibaryon rapidity density  yields  are studied,
relative to, and as function of,  participant number, and produced hadron
yields.  
\end{abstract}

 \maketitle
The experiments at the Relativistic Heavy Ion Collider (RHIC) 
study  the properties of quark-gluon plasma (QGP) matter, a state that was  present in
the early Universe. We  explore in this paper the  production of  baryons and 
antibaryons containing one or more strange quarks. Strange antibaryons 
offer a unique signature of the presence of QGP~\cite{antibaryon}. Their yields 
probe  how QGP  turns into conventional matter (hadronization).

At origin of this strange (anti)baryon signature is the high abundance of 
strangeness expected in QGP, combined with  the high matter density at 
hadronization of this entropy rich phase~\cite{Letessier:1992xd}. 
A relatively small  conventional  production
background enhances the importance of (multi)strange antibaryon
QGP  signature~\cite{Weber:2002pk}. We use
results of  an analysis of the general pattern of particle 
production at central rapidity as function of  $A$~\cite{Rafelski:2004dp}
for RHIC $\sqrt{s_{\rm NN}}=200$ GeV Au--Au reactions. 
These result were obtained within the statistical hadronization model~\cite{share} 
(SHM)   description of the PHENIX experimental
 central rapidity yields~\cite{Adler:2003cb}
 $dN_i/dy$, $N_i=\pi^\pm,{\rm K}^\pm, p$ and $\bar p$ 
along  with STAR experiment relative yields 
${\rm K}^*(892)/{\rm K}^-$~\cite{Zhang:2004rj} 
and  $\phi/{\rm K}^-$~\cite{Adams:2004ux}.  

The hadronization conditions for strangeness are presented for 
the first time here and will be used to predict the centrality dependent
yields of strange and multi-strange baryons and antibaryons. These results
are obtained without any adjustment or revision of  Ref.~\cite{Rafelski:2004dp}.
The statistical model parameters allowing us to differentiate the degree of 
chemical equilibration is the phase space occupancy $\gamma_i$. 
$\gamma_i$ has  for each flavor `$i$' of quarks a similar role as the 
``blackness'' of a gray-body  photon-emitter: for $\gamma<1$ there are fewer particles
than expected from a black-body source, and for $\gamma>1$ more. 
The   SHM  analysis of the final state hadron yields  presumes thermal equilibrium, 
and  we consider  all   chemical (non-)equilibrium conditions studied in literature:

1) Chemical non-equilibrium of  valence light and strange
quark pair abundance; this we expect to  
arise in  a rapid transformation of the deconfined quark-matter 
into free-streaming hadrons. In such a sudden hadronization, there is 
no time to re-equilibrate the final state yields determined by 
fragmentation and recombination of available QGP partons. 
Because the (chemically equilibrated) phase space in QGP, 
in  general, has a different density compared to the
hadronic gas (HG), QGP chemical  equilibrium described by quark-matter 
space occupancies  $\gamma_s^{\rm Q}=1,\ \gamma_q^{\rm Q}=1$ 
implies hadron  chemical non-equilibrium, with the parameters  
 $\gamma_s^{\rm H}\ne 1,\ \gamma_q^{\rm H}\ne 1$. 

2) Chemical semi-equilibrium, which may arise for sufficiently slow
hadronization,  allowing to re-equilibrate the light quark 
abundances after hadron formation, $ \gamma_q^{\rm H}\to  1$.
Since strangeness re-equilibration is a considerably slower reaction
process~\cite{Koch:1986ud}, it  did not have sufficient time to develop. 
Alternatively, this chemical semi-equilibrium  can  also arise in
reactions not involving a phase change, with particle production
in hadron sector evolving to chemical equilibrium for non-strange 
hadrons,  but strange hadrons  not having enough time to reach 
the strangeness yield `absolute' chemical equilibrium. 

3) Full chemical equilibrium,  
$\gamma_s^{\rm H}=\gamma_q^{\rm H}=1$, requires 
a complete re-equilibration, or, if there is no phase change, equilibration
of all types of hadronic particles. 

We assess the viability of QGP strange (anti)baryon signatures of deconfinement
by comparing these three scenarios as function of $A$. 
 For very peripheral collisions,  also studied  in Ref.\,\cite{Rafelski:2004dp}, 
there was a significant deviation from chemical equilibrium with 
the strangeness yield being significantly reduced, as measured both, by  quark 
occupancy $\gamma_s^{\rm H}/\gamma_q^{\rm H}<1$, and 
the (nearly conserved) strangeness to entropy ratio $s/S\simeq 0.019$. 
Moreover,  in this limit, the 
data fit employing   chemical equilibrium $\gamma_s^{\rm H}=\gamma_q^{\rm H}=1$
did not yield a   confidence level $P[\%]$ comparable to the chemical
non-equilibrium approach. The present analysis includes all results for $A>20$
which yield  $P[\%]>60\%$. With increasing $A$, for the 
chemical non-equilibrium statistical hadronization model, the 
strangeness occupancy $\gamma_s^{\rm H}$ was rising rapidly, reaching for most
central collisions the value $\gamma_s^{\rm H}=2.4$\,. The relative 
yield of strangeness per entropy was found for the
 most central collisions, $A\simeq 350$,
at $s/S\simeq 0.028\pm0.002$, which corresponds to the expectations 
one has for nearly chemically equilibrated   QGP.

\begin{figure}[thb]
\vskip -0.5cm
\psfig{width=8cm,height=11.8cm,figure=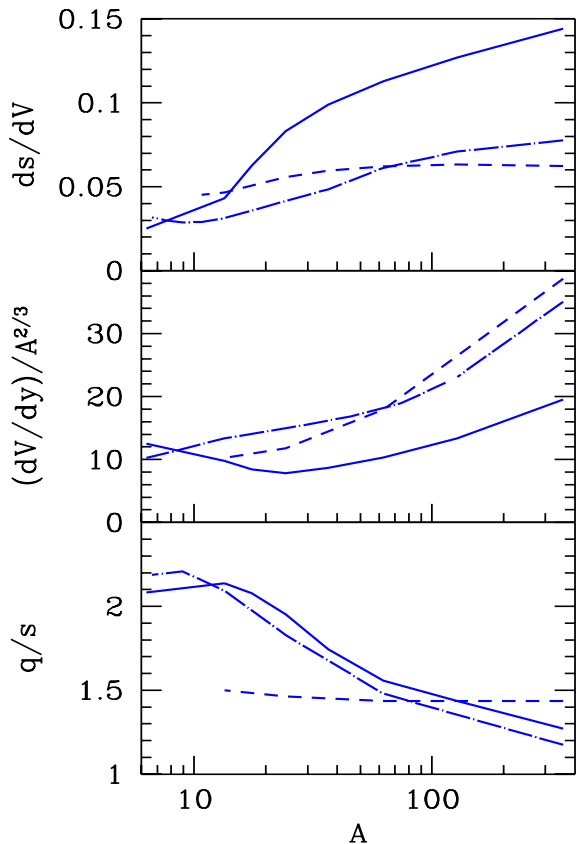 }
\vspace*{-0.6cm}
\caption{\label{HadsCon}
Top panel:  Strangeness density 
$ds/dV\, [{\rm fm}^{-3}]$; middle panel: geometrically 
reduced hadronization volume 
$dV/dy /A^{2/3}\, [{\rm fm}^{-3}]$;  and  bottom panel:
the ratio of light quark to strange quark density  $q/s$
as function of  $A$.
 Solid lines: chemical non-equilibrium; chain lines: chemical
semi-equilibrium; and dashed lines: chemical equilibrium.  
}
\vskip -0.5cm
\end{figure}
 
 We show, in the top panel of  Fig.\,\ref{HadsCon},  the average  strangeness 
density of the hadron source at central rapidity, as function of   $A$,
 at the time of hadronization:
$(ds/dy)/ (dV/dy)={ds/dV}=  {d\bar s/dV}$.
For very  peripheral collision systems, the central rapidity density is at 
$ds/dV= 0.03/$fm$^3$. As $A$ increases, there is a significant 
increase in the strangeness density at hadronization in the chemical
non-equilibrium model, with the full non-equilibrium (solid line) approaching
a five-fold increase.  $ds/dV$  doubles in the semi-equilibrium case 
(chain line) and increases by 50\% in chemical equilibrium model (dashed line).

This result combines two multiplicative factors:\\ 1) The strangeness
yield per unit of rapidity, $ds/dy$, which, in principle,
can be directly  estimated from the experimental yields of particles which
were fitted, and is thus model independent.\\   
2) The statistical volume 
per rapidity $dV/dy$, which is a model dependent 
result of the fit to the  global particle yield
pattern.\\ 
We show its behavior in the middle panel of  Fig.\,\ref{HadsCon}, reduced
by the initial state geometric scale factor $A^{2/3}$. For small $A$,
the chemical non-equilibrium result shows a 40\% reduction of the 
residual volume with increasing $A$, as would be expected for an 
increasing stopping power, {\it e.g.\/}, due to the progressive 
onset of deconfinement. 

All models show, for large $A$, an increase  
of $(dV/dy) /A^{2/3}$   with $A$, which can be interpreted as being due to 
the grater dynamic expansion prior to hadronization of the larger, 
initially  more compressed system. For the chemical non-equilibrium 
model,  $(dV/dy) /A^{2/3}$ increases by 180\% compared to minimum near $A=25$. 
The  yet larger increases, by a factor 3.5--4, 
 seen for the  the chemical
semi-equilibrium and equilibrium  models, compensate  for the chemical
non-equilibrium  factor $\gamma_q^{\rm H} \simeq 1.6$. Some may consider such an 
increase excessive, in which case the result  is an indirect evidence 
for   $\gamma_q^{\rm H} >1$. 

In order to gauge the impact of the strangeness hadronization
density increase, we have to  compare it to light quark density.
as this  effectively eliminated the model dependence.
To obtain  such a $q/s$ ratio, we  estimate  the light quarks number, 
including any originating in  hadronization fragmentation, 
from the final state entropy content. For quarks with thermal 
mass $m_q\simeq 2T$,  for the classical massive gas 
the  entropy per particle is $S/N\simeq 5$. 
Thus: 
\begin{equation}
{q\over s}\simeq \frac1{ds/dy}\left[{1\over k (S/N)} 
        \left({dS\over dy}- k_s \frac{S}{N} {ds\over dy}\right)\right].
\end{equation}
 We reduced the total entropy content by the two $(k_s=2)$ fractions
$s$ and $\bar s$, and 
divided by   $k (S/N)=20$ to obtain the approximate average yield of the $k=4$ 
components $q=u,\,\bar u,\, d$ and $ \bar d$. 

The chemical non-equilibrium $q/s$ (solid line in 
bottom panel of Fig.\,\ref{HadsCon}) and semi-equilibrium  (chain line)
ratios  $q/s$ are identical within the error margin of this estimate.  
Strangeness is, at small $A$, suppressed by more than factor 2.
For $A\to 350$,   a 20\% excess of  light quark number compared
to strange quark number  remains.  This    can be
understood, {\it e.g.\/},  as  being
 due to the thermal strange quark mass being slightly
larger than the light quark thermal mass, and both light and strange
quarks approaching chemical
equilibrium yields  in the QGP. The hadron-phase chemical equilibrium 
result (dashed line) shows no variation with $A$. 
The increased yield
of strangeness is consistent with the rise of the reduced reaction volume, 
which is implying an increase in the lifespan of the dense phase.

In the following, we  discuss results which 
 include  decay products of weak decays: we count 
all pions from  ${\rm K}_{\rm S}$ decay, we include  10\% ${\rm K}_{\rm L}$ 
decay products, and  all hyperon 
decay, {\it i.e.\/},  both pions and baryons are accepted. In particular, 
we included in our $d\Lambda/dy$ and $d\overline\Lambda/dy$, all weak decays 
of  $d\Xi^-\!\!/dy,\,d\overline{\Xi}^+\!\!/dy$ and $d\Omega^-\!\!/dy,\,
d\overline{\Omega}^+\!\!/dy$.
 These   idealized
assumptions will cause  minor calculable correction 
 for more realistic  experimental acceptances. 
The interested reader
can use the SHM implementation~\cite{share} to obtain results for
other weak decay  conditions. 

\begin{figure}[thb]
\hspace*{-.3cm}
\psfig{width=8.3cm,height=10.5cm,figure= 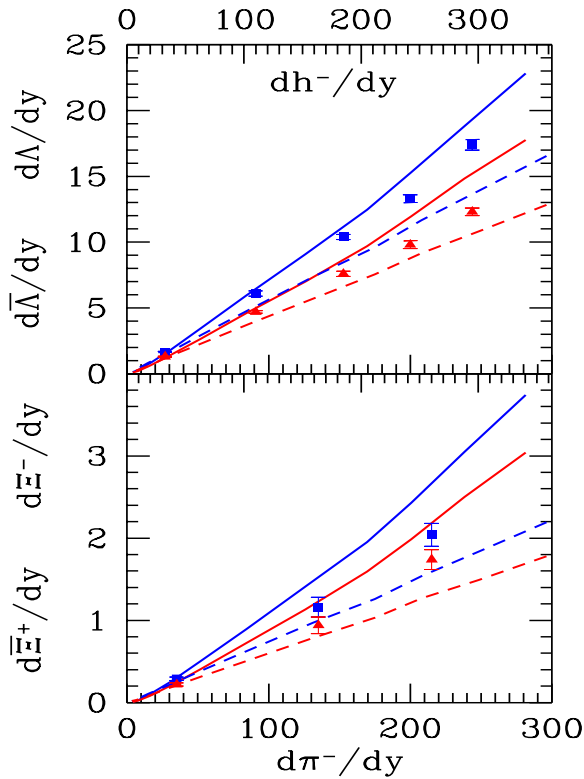}
\vspace*{-0.3cm}
\caption{\label{HypYields}
(color online)  The predicted    hyperon (blue)
and, in same panel directly below,  
antihyperon (red) yields.  
The rapidity yields are shown as function
of both $d\pi^-\!\!/dy$ (bottom scale) and $dh^-\!\!/dy$ (top scale)  
 for $\sqrt{s_{\rm NN}}=200$ GeV Au--Au reactions.
Solid lines for chemical 
non-equilibrium, and dashed lines for the chemical equilibrium.
 The experimental results (square symbols
 for hyperons and triangles for antihyperons)
 are from  $\sqrt{s_{\rm NN}}=130$ GeV Au--Au reactions, presented
here for reader orientation.
 }
\vskip -0.3cm
\end{figure}

We present, in Fig.\,\ref{HypYields}, the predicted production rate
of $d\Lambda/dy,\,d\overline\Lambda/dy$ (top panel), 
 $d\Xi^-\!\!/dy,\, d\overline\Xi^+\!\!/dy$ (bottom panel) as function 
of $dh^-\!\!/dy$ and $d\pi^-\!\!/dy$ at central rapidity. 
 The dependence on $h^-$ is indicated in top, and on $\pi^-$ in bottom of
the figure. We note that the equilibrium model (dashed lines) predicts
fewer (anti)hyperons. We omitted the chemical semi-equilibrium lines which are
in-between the two other models. 
The available experimental results shown 
are for $\sqrt{s_{\rm NN}}=130$ GeV Au--Au 
rather than here considered 200 GeV reactions,
$d\Lambda/dy,\, d\overline\Lambda/dy$~\cite{Adler:2002uv}, and 
 $d\Xi^-\!\!/dy,\, d\overline\Xi^+\!\!/dy$~\cite{Adams:2003fy}. 
Considering that  at $\sqrt{s_{\rm NN}}=200$ GeV the yields should 
be higher, the  chemical nonequilibrium model is more likely
to be consistent with the 200 GeV data. 

\begin{figure}[htb]
\psfig{width=8.3cm,height=11cm,figure=  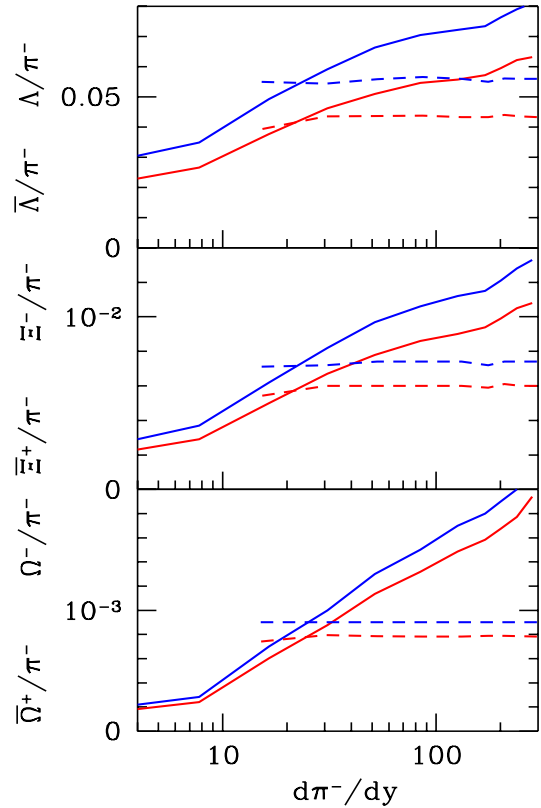}
\vspace*{-0.3cm}
\caption{\label{STARhyppi}
(color online)  The yields of  hyperons (blue) 
and antihyperons (red),
$\Lambda,\ \overline\Lambda$ (top panel), 
 $\Xi^-\!\!,\ \overline\Xi^+$ (middle panel),
and $\Omega^-\!\!,\ \overline\Omega^+$ bottom normalized
with $\pi^-$ yield, as function of $d\pi^-\!\!/dy$ yield.
The lines are as in Fig.\,\ref{HypYields}. 
 }
\vskip -0.3cm
\end{figure}

We now turn to discuss the enhancement effect in production of 
hyperons and antihyperons. A way to study this enhancement 
is to consider the yield of the particle 
 per hadron and/or per pion as function of 
hadron/pion yield. 
We   present, in Fig.\,\ref{STARhyppi}, the full set of hyperon
 theoretical predictions    normalizing these with the
$\pi^-$ yield.  We see that the normalized
chemical equilibrium yield is  flat, while the chemical non-equilibrium
yields predict enhancement with $A$. This  
effect can be  seen, in Fig.\,\ref{HypYields}, as a slight up-bend 
in the particle yield at high $A$.

Such an up-bend  is visible in the 
experimental $\Lambda$ and $\overline\Lambda$ results for 
$\sqrt{s_{\rm NN}}=130$ GeV Au--Au reactions~\cite{Adler:2002uv}, however
the data was not analyzed to reveal this effect. Considering 
result shown in Fig.\,\ref{STARhyppi}, we  suggest
 that a fit to the experimental centrality 
dependence should be made allowing for a non-linear shape, {\it e.g.\/}:
\begin{equation}\label{hfit}
dN/dy=a_1\,h^-+0.5 a_2\,(h^-)^2,
\end{equation}
 where predicted $a_2$ is, up to a factor $h^-\!\!/\pi^-=1.2$, 
the slope  seen in   Fig.\,\ref{STARhyppi}, solid  line. 

Should  $a_2\to 0$, this would constitute reliable evidence for validity of 
the chemical equilibrium model: such a result relates closely
to the flat $q/s$ relative  yield, in Fig.\,\ref{HadsCon},
bottom panel. $a_2\to 0$  means that 
there is no centrality dependence in relative strange and light 
quark abundances, and densities,  at hadronization.
This is  expected for chemical equilibrium, but
would require an extraordinary  coincidence to occur  at several  
centralities for chemical non-equilibrium. Therefore, this  is a   
test for chemical equilibrium, more appropriate   than the more common  
study of $\gamma_s^{\rm H}\to 1$ and  $\gamma_q^{\rm H}\to 1$. In the
latter case, the fitted value of $\gamma_{s,q}^{\rm H}$  can
be different from unity, due to incompleteness of the SHM set of 
hadron states.

\begin{figure}[thb]
\psfig{width=8.5cm,figure= 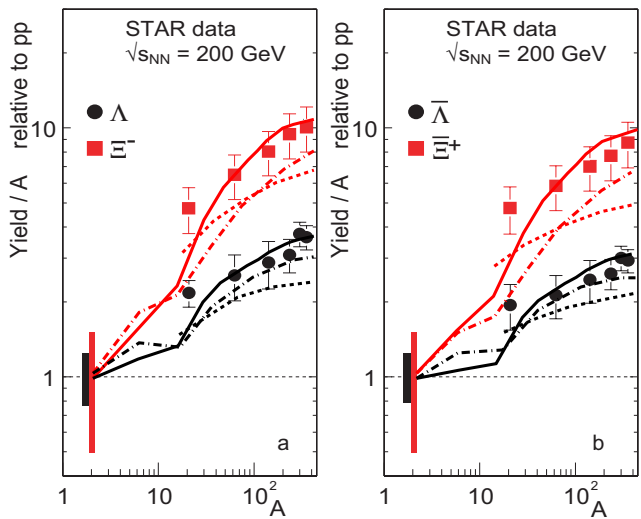}
\vspace*{-0.6cm}
\caption{\label{STARhyp}
(color online) $\sqrt{s_{\rm NN}}=200$ GeV yields  
of hyperons $d\Lambda/dy$ and $d\Xi^-\!\!/dy$,
on left, and antihyperons
 $d\overline\Lambda/dy$ and $d\overline\Xi^+\!\!/dy$,
on right, normalized with, and as function of, 
$A$, relative to these yields in $pp$ 
reactions. The lines are as in Fig.\,\ref{HypYields}. 
}
\vskip -0.3cm
\end{figure}

There is another  way to study enhanced production of strange 
hyperons~\cite{Andersen:1999ym}: we consider   the hyperon yield
normalized by $A$,  as function of $A$. To obtain enhancement, we divide  
by the  background signal defined to be the experimental yield  for the 
smallest available system, {\it e.g.\/},  $A=2$.
The STAR collaboration has stated the central rapidity
yield of $d\Lambda/dy,\, d\overline\Lambda/dy,\, d\Xi^-\!\!/dy$ 
and $d\overline\Xi^+\!\!/dy$, at $\sqrt{s_{\rm NN}}=200$ GeV in this way, 
see  Fig.\,\ref{STARhyp}~\cite{Caines:2004ej}. 
The experimental  reduced yields vary for large $A$,  with $A$. This 
behavior cannot be due to  a reinterpretation of  the enhancement as being due to 
the $pp$ yield suppression, for it occurs at relatively large $A$.

We overlay, over these data, the {\it unsmoothed} results we obtained with SHM. 
To normalize the SHM predictions, we use the central values of
$pp$ yields stated in Ref.\,\cite{Heinz:2005vu}:
$d(\Lambda+\overline\Lambda)/dy=0.066\pm 0.006  $,
$d(\Xi^-\!\!+\overline{\Xi}^+\!)/dy= 0.0036\pm 0.0012 $,
$\overline\Lambda/\Lambda= 0.88\pm 0.09$ and
$\overline{\Xi}^+\!\!/\Xi^-\!=0.90\pm 0.09 $. Thus, we also 
extend the lines from $A=6.3$, the smallest  value considered in peripheral 
Au--Au interactions, to  $A=2$, where they assume by definition the value unity.
Since the chemical equilibrium SHM fit is not valid  below $A=20$ we do not 
continue this line.  The edge, near to $A=20$, is
signaling the change in behavior of the SHM fit associated with a
 possible phase change~\cite{Rafelski:2004dp,Letessier:2005qe}.

The agreement of the chemical non-equilibrium result
with all four data sets is remarkable. The reader should note
that the relative yields and $A$-dependence of the 
$\Lambda,\overline\Lambda,\Xi,\overline\Xi$ are correctly
described by SHM model parameters derived from study of other 
particles. This is a remarkable accomplishment of the SHM. 
This remark applies to all three models considered, considering the magnitude of the
normalization uncertainty, indicated at $A=2$, in Fig.\,\ref{STARhyp}
--- here we note 
that we cannot with required
precision extract the absolute yields from 
Fig.\,\ref{STARhyp}~\cite{Caines:2004ej}, and
enter these in Fig.\,\ref{STARhyppi}, considering
the double logarithmic representation.

In this paper, we have explored the strangeness content of the RHIC fireball 
at its hadronization point  found in Ref.\,\cite{Rafelski:2004dp}.
We showed, as function of $A$, the behavior of the relative pre-hadron formation 
yields of light and strange quarks. Within the context of chemical non-equilibrium,
the  strangeness content increases rapidly with $A$. 
The reaction volume scales  with $A^{2/3}$, however, the residual
variation is characteristic of  phase properties and expansion dynamics.
 
The   increase in $s/q$  with $A$, Fig.\,\ref{HadsCon}, is 
 imaged  in the rise of the specific per hadron yields of  
strange antibaryons, Fig.\,\ref{STARhyppi}, with centrality.  
The variation of the strangeness yield with the reaction volume
(impact parameter) is furthermore
the source of the rapid rise of  multi-strange (anti)baryon
yield enhancement seen in  Fig.\,\ref{STARhyp}. 

The absolute strange (anti)baryon rapidity yields, Fig.\,\ref{HypYields},
predicted for $\sqrt{s_{\rm NN}}=200$ GeV, distinguish the hadronization 
models: the yields of $d\Xi^-\!\!/dy$ and $d\overline{\Xi}^+\!\!/dy$ 
are, for the chemical non-equilibrium case, 80\% above those
expected in the chemical equilibrium approach.
The high experimental hyperon yields reported at   $\sqrt{s_{\rm NN}}=130$ GeV   
favor the chemical non-equilibrium 
hadronization.  Future experimental results can  as we have shown
provide decisive  distinction between the hadronization pictures considered. 

Work supported by a grant from: the U.S. Department of
Energy  DE-FG02-04ER4131.
LPTHE, Univ.\,Paris 6 et 7 is: Unit\'e mixte de Recherche du CNRS, UMR7589.
 

\vspace*{-0.3cm}

\end{document}